\documentclass[aps,pra,groupedaddress]{revtex4-2}
\usepackage{enumitem}
\usepackage{xcolor}
\usepackage{soul}
\usepackage{tikz}
\usetikzlibrary{quantikz}

\usepackage{empheq}

\usepackage{mathtools,amsfonts,amsmath,amssymb,mathrsfs}
\usepackage{calrsfs}
\usepackage{exscale}
\usepackage{xfrac}
\usepackage{stmaryrd}
\usepackage{eucal}
\usepackage{graphicx} 
\usepackage{subfig}
\usepackage{graphics}
\usepackage{epsfig}
\usepackage{graphicx,color}           
\usepackage{color}
\usepackage{dcolumn}
\usepackage{bm}
\usepackage{float}
\usepackage{hyperref}
\usepackage[normalem]{ulem}  
\usepackage{multirow}  
\hypersetup{colorlinks,linkcolor={blue},citecolor={red},urlcolor={red}}
\graphicspath{{./figs/}}
\newlength{\subcolumnwidth}

\newcommand{\nextsubcolumn}[1][]{%
  \cr\noalign{\hfill}
  \if\relax\detokenize{#1}\relax\else\hsize=#1\setlength{\subcolumnwidth}{\hsize}\fi
}

\makeatletter
\let\p@section\@empty
\let\p@subsection\@empty
\let\p@subsubsection\@empty
\makeatother

\begin{document}

%
%
\title{Atomic non-classicality: A study of the anti-Jaynes-Cummings interaction}
%
%
\author{Christopher Mayero}
\email[E-mail address: ]{cmayero@tmu.ac.ke}
\affiliation {Tom Mboya University, Department of Physics, C19,  P.O. Box 199-40300, Homabay, Kenya}
\date{\today} 
%
%
\begin{abstract}
We apply the Wigner-Yanase skew information, $I(\hat{\rho}, \hat{K})$, as  a quantum information quantifier of atomic non-classicality in the dynamics generated by the anti-Jaynes-Cummings (AJC) Hamiltonian when a two-level atom in an initial atomic ground state $\mid g\rangle$ couples to a single mode of squeezed coherent light. We investigate the effect of variation of squeeze parameter, $r$, field intensity, $\mid\alpha\mid^2$, and coupling strength parameter, $\xi$, on the dynamics of, $I(\hat{\rho}, \hat{K})$. We observe that, $I(\hat{\rho}, \hat{K})$ records mixed state values for all variations of $r,\mid\alpha\mid^2,~\xi$ congruent with squeezing effects.\\
   
\end{abstract}

\maketitle
%
%
\section{Introduction}
\label{sec:introduction}

The simplest single-mode spin-boson model that describes the interaction between light and matter is the quantum Rabi model (QRM) \cite{rabi1936process,rabi1937space,braak2011integrability,
braak2016semi,xie2017quantum}. This Hamiltonian has two dynamical frames \cite{omolo2021anti}; the rotating frame (RF) and the counter (anti)-rotating frame (CRF). The dynamics in the rotating frame \cite{mayero2021rabi,cmayero2021pol} is governed by the Jaynes-Cummings (JC) interaction process and through a $U(1)$ symmetry transformation generated by the JC excitation number operator $\hat{N}=\hat{a}^{\dagger}\hat{a}+\hat{\sigma}_+\hat{\sigma}_-$. In this case, the QRM is approximated by an effective JC Hamiltonian, $\hat{H}$, \cite{jaynes1963comparison,shore1993jaynes,shore_knight_1993, omolo2017conserved,gerry2005introductory,xie2017quantum,
allen1987optical,haroche2013nobel} in the rotating wave approximation (RWA). In the RWA, the coupling strength, $\lambda$, is much weaker than the mode frequency, $\omega$, i.e., $\lambda<\mid\omega\mid$. In this circumstances, if the qubit is close to resonance, $\mid\omega_0\mid-\mid\omega\mid\simeq 0$, and $\mid\omega_0+\omega\mid\geq\mid\omega_0-\omega\mid$ holds, the RWA can be applied. This implies neglecting terms that rotate at frequency $\omega_0+\omega$, leading to the JC Hamiltonian. On the other hand, dynamics in the CRF, is controlled by the anti-Jaynes-Cummings (AJC) interaction process \cite{omolo2021conserved,omolo2021anti,mayero2021rabi,
mayero2023photon,mayero2024squeezed} through a $U(1)$ symmetry transformation generated by the AJC conserved excitation number operator, $\hat{\overline{N}}=\hat{a}\hat{a}^{\dagger}+\hat{\sigma}_-\hat{\sigma}_+$. The QRM in this sense is approximated by an effective AJC Hamiltonian, $\hat{\overline{H}}$, in the counter-rotating wave approximation (CRWA) in the form $\mid\omega_0-\omega\mid\geq\mid\omega_0+\omega\mid$. Here, terms that rotate at frequencies $\omega_0-\omega$ are dropped resulting to the AJC Hamiltonian, $\hat{\overline{H}}$. The AJC model, $\hat{\overline{H}}$, as a component of the QRM in the CRF, is exactly solvable \cite{omolo2021anti} and features a conserved excitation number operator. Recently, It has been demonstrated \cite{cmayero2021pol,mayero2021rabi} that it is feasible to generate entangled anti-symmetric atom-field states, entangled symmetric atom-field states in the AJC,\,JC processes respectively while in \cite{mayero2023photon,mayero2024squeezed} non-classicality in the AJC process as measured by the Mandel-Q parameter is discussed. During the AJC interaction \cite{mayero2021rabi}, the Rabi oscillations occur in the reverse sense in relation to that during the JC interaction mechanism.

In this article, we study atomic non-classicality dynamics  generated by the AJC Hamiltonian when a two-level atom interacts with a quantised field mode initially in a squeezed state \cite{gerry2005introductory,Loudon1987Squeezed,Knight2004,
mandel1986non,zaheer1990advances,moya1992interaction,
milburn1984interaction,satyanarayana1989ringing,
schleich1987oscillations,schleich1988area,teich1989squeezed}, measured by the Weigner-Yanase skew information \cite{wigner1963information}.

The Wigner-Yanase skew information \cite{wigner1963information}  

\begin{eqnarray}
I(\hat{\rho},\hat{K})=-\frac{1}{2}tr\left[\sqrt{\hat{\rho}},\hat{K}\right]\nonumber
\label{eq:wigner0}
\end{eqnarray}
is a measure of the information content of a pure state or a mixed state, $\hat{\rho}$, skew to an observable (Hermitian operator), $\hat{K}$, where $\left[\,,\,\right]$ denotes the commutator. In the following, without loss of generality, we shall interpret the time evolution of the Wigner-Yanase skew information based on the simplified redefinition in \cite{dai2019information} christened atomic spin non-classicality quantifier in the form

\begin{eqnarray}
N(\hat{\rho})=\frac{1}{2}\left(1-\sqrt{1-\mid\vec{r}\mid^2}\right)\nonumber
\end{eqnarray}
where $\vec{r}=r_x\hat{i}+r_y\hat{j}+r_z\hat{k}$ is the Bloch vector. It therefore follows that; when $\mid\vec{r}\mid=0$, $N(\hat{\rho})=0$ corresponds to a maximally mixed (maximally entangled) state, $\mid\vec{r}\mid=1$, $N(\hat{\rho})=\frac{1}{2}$  pure (product) state else $0<\mid\vec{r}\mid<1$, $0<N(\hat{\rho})<\frac{1}{2}$, mixed (entangled) state.

It has been established in \cite{mayero2024squeezed,mayero2023photon}  that a non-vanishing residual detuning, $2\omega$,  exists in the AJC process and absent in the standard JC resonance, $\delta=0$. This being a fundamental distinctive internal dynamical property of the AJC interaction mechanism, a residual detuning parameter, $\xi=\frac{\omega}{\lambda}$ present in the general definition of sum frequency, $\overline{\delta}=2\xi\lambda$ at $\delta=0$ is interpreted as a coupling strength, where $\omega,\,\lambda$ are field mode frequency, coupling constant respectively. Small values of $\xi$ such that $\lambda\gg\omega$ signifies strong atom-field AJC interaction. We study the effect of the coupling strength parameter on the dynamics of Wigner-Yanase skew information during the atom-field AJC interaction process in addition to effects arising from variation of field intensity, $\mid\alpha\mid^2$  and squeeze parameter, $r$.  

This work is organised as follows; Sec.~\ref{sec:method} introduces the AJC theoretical model and its time evolution; in Sec.~\ref{sec:dynas} non-classicality as measured by the Weigner-Yanase skew information is provided and Sec.~\ref{sec:conc} is the conclusion.

\section{The AJC model and its time evolution}
\label{sec:method}

The AJC Hamiltonian as a component of the QRM in the CRF determined in anti-normal order form is defined as
\begin{eqnarray}
\hat{\overline{H}}=\hbar\left[\omega\hat{\overline{N}}+\overline{\delta}\hat{s}_z+\lambda(\hat{a}\hat{s}_-+\hat{a}^\dagger\hat{s}_+)-\frac{1}{2}\omega\right]\quad;\quad\hat{\overline{N}}=\hat{a}\hat{a}^\dagger+\hat{s}_-\hat{s}_+\quad;\quad\overline{\delta}=\omega_0+\omega\quad;\quad\lambda=2g.
\label{eq:CRF1}
\end{eqnarray}
Here, $\omega$ is the field mode frequency; $\omega_0$ is the atomic transition frequency; $g$ is the atom-field coupling constant; $\hat{a},\,\hat{a}^{\dagger}$ are the field annihilation and creation operators, respectively obeying $\left[\hat{a},\hat{a}^{\dagger}\right]=1$; $\hat{s}_z$ is the atomic inversion operator; and $\hat{s}_+=\hat{s}_x+i\hat{s}_y,\,\hat{s}_-=\hat{s}_x-i\hat{s}_y$ are the atomic transition operators and satisfy $\left[\hat{s}_+,\hat{s}_-\right]=2\hat{s}_z,\,\left[\hat{s}_z,\hat{s}_{\pm}\right]=\pm \hat{s}_{\pm}$ with $\hat{s}_x=\frac{1}{2}\hat{\sigma}_x,\,\hat{s}_y=\frac{1}{2}\hat{\sigma}_y,\,\hat{s}_z=\frac{1}{2}\hat{\sigma}_z$ the well-known Pauli spin operator matrices. The operator, $\hat{\overline{N}}$, is the AJC conserved excitation number operator.    

We express the AJC sum frequency detuning, $\overline{\delta}=\omega_0+\omega$, in Eq.~\eqref{eq:CRF1} in terms of the JC difference frequency detuning, $\delta=\omega_0-\omega$, in the form
\begin{eqnarray}
\overline{\delta}=\delta+2\omega
\label{eq:ajcsumf}
\end{eqnarray}
to  conveniently vary $\delta$, AJC residual detuning, $2\omega$, separately during analysis of atomic non-classicality dynamics generated by the AJC Hamiltonian defined in Eq.~\eqref{eq:CRF1}. With reference to Eq.\eqref{eq:ajcsumf}, notice that when $\delta=0$, $\overline{\delta}=2\omega$ is a non-vanishing residual frequency detuning observed only in the AJC process and defined earlier in  \cite{forn2010observation,pradhan2004effects,cardoso2005situ} as the Bloch-Siegert oscillation (BSO).
 
In this work, we consider an atom initially in an atomic ground state $\mid g\rangle$ 
and the field mode initially in a squeezed coherent state $\mid \alpha,r\rangle_{t=0}$ defined as \cite{gerry2005introductory} 
\
\begin{eqnarray}
\mid \alpha,r\rangle_{t=0}&=&S_n\mid n\rangle \quad;\quad\nonumber\\
S_n&=&\frac{1}{\sqrt{\cosh(r)}}\exp\left[-\frac{1}{2}|\alpha|^2-\frac{1}{2}\alpha^{*2}\tanh(r)\right]\nonumber\\
&\times&\sum_{n=0}^\infty\frac{\left[\frac{1}{2}\tanh(r)\right]^{\frac{n}{2}}}{\sqrt{n!}}\times H_n\left[\left(\alpha\cosh(r)+\alpha^{*}\sinh(r)\right)\left(\sinh(2r)\right)^{-\frac{1}{2}}\right]\quad;\quad\theta=0\nonumber\\&&
\label{eq:sqc}
\end{eqnarray}
where $\alpha$ is the coherent state mean photon number amplitude, $r$ is the squeeze parameter and $S_n$ is the probability amplitude for \texttt{n}-photons in the state $\mid n\rangle$, giving the probability of finding \texttt{n} photons in the field in the form \cite{gerry2005introductory}
\begin{eqnarray}
P(n)&=&\mid S_n\mid^2=\mid\langle n\mid \alpha,r\rangle\mid^2\nonumber\\
&=&\frac{\left[\frac{1}{2}\tanh(r)\right]^n}{n!\cosh(r)}\exp\left[-\mid\alpha\mid^2-\frac{1}{2}\left(\alpha^{*2}+\alpha^2\right)\tanh(r)\right]\nonumber\\
&\times&\Big\vert H_n\left[(\alpha\cosh(r)+\alpha^*\sinh(r))(\sinh(2r))^{-\frac{1}{2}}\right]\Big\vert^2~.
\label{eq:photonno}
\end{eqnarray}

The initial average photon number, $\langle\hat{n}\rangle$, of a squeezed coherent state is a sum of the coherent and the squeeze contributions, which in this respect

\begin{eqnarray}
\langle\hat{n}\rangle=\mid\alpha\mid^2+\sinh^2(r)\quad;\quad \mid\alpha\mid^2=\langle\hat{a}\hat{a}^{\dagger}\rangle_{t=0}=\langle\hat{a}^{\dagger}\hat{a}+1\rangle_{t=0}.
\label{eq:pinitial}
\end{eqnarray}

The composite atom-field mode initial state $\mid\psi_{gr\alpha}\rangle$  takes the form \cite{scully1997quantum}
\begin{eqnarray}
\mid\psi_{gr\alpha}\rangle&=&\mid g\rangle\otimes\mid\alpha,r\rangle
=\sum_{n=0}^\infty S_n~\mid g,n\rangle~.
\label{eq:initial}
\end{eqnarray}

Acting on this initial state, Eq.~\eqref{eq:initial}, the AJC Hamiltonian in Eq.~\eqref{eq:CRF1}, generates a time evolving state vector,  $\mid\overline{\Psi}_{gr\alpha}(t)\rangle$, expressed in Schmidt decomposition \cite{gerry2005introductory,jaeger2009entanglement,nielsen2011quantum} form  

\begin{eqnarray}
\quad\quad\quad\mid\overline{\Psi}_{gr\alpha}(t)\rangle&=&e^{-\frac{i}{\hbar}\hat{\overline{H}}t}~\mid\psi_{gr\alpha}\rangle=\sum_{n=0}^\infty\Big[e^{-i\omega \left(n+1\right)t}~S_n~\Big(\cos(\overline{R}_{gn}t)\nonumber\\&&+i\overline{c}_{gn}\sin(\overline{R}_{gn}t)\Big)\mid g\rangle-ie^{-i\omega nt}~S_{n-1}~\overline{s}_{gn-1}\sin(\overline{R}_{gn-1})\mid e\rangle\Big]\otimes\mid n\rangle~;\nonumber\\
\overline{R}_{gn}&=&\frac{\lambda}{2}\sqrt{4n+4+(\beta+2\xi)^2}\quad;\quad\overline{c}_{gn}=\frac{(\beta+2\xi)}{\sqrt{4n+4+(\beta+2\xi)^2}}\nonumber\\
\overline{s}_{gn}&=&\sqrt{\frac{4(n+1)}{4n+4+(\beta+2\xi)^2}}\quad;\quad\overline{\delta}=(\beta+2\xi)\lambda\quad;\quad\xi=\frac{\omega}{\lambda}\quad;\quad\beta=\frac{\delta}{\lambda}~.
\label{eq:ajcev}
\end{eqnarray}
In Eq.~\eqref{eq:ajcev}, $\beta=\frac{\delta}{\lambda},\,\xi=\frac{\omega}{\lambda}$ are dimensionless detuning parameters, with $\xi$ a non-vanishing dimensionless frequency detuning parameter interpreted as a coupling strength such that small values of $\xi$ in the range $(0,1)$ specify a strongly coupled atom-field system in the AJC interaction and large values of $\xi$ greater than unity characterises a weakly coupled atom-field quantum system in AJC process.


\section{Dynamics of Atomic Non-classicality}
\label{sec:dynas}
In this section we apply the Wigner-Yanase skew information \cite{wigner1963information} as a measure of the information content of the time-evolving reduced density operator of the atom $\hat{\overline{\rho}}_a^g(t)$ skew to a Hermitian operator, $\hat{s}_z$, in the form
\begin{subequations}
\begin{eqnarray}
I\left(\hat{\overline{\rho}}_a^g(t),\,\hat{s}_z\right)=-\frac{1}{2}tr\left[\sqrt{\hat{\overline{\rho}}_a^g(t)}~,~\hat{s}_z\right]^2\quad;\quad \hat{K}=\hat{s}_z.
\label{eq:wigner1}
\end{eqnarray} 

The skew information in Eq.~\eqref{eq:wigner1}, constitutes an alternative measure for the information content of the time-evolving state $\hat{\overline{\rho}}_a^g(t)$ skew to the observable, $\hat{s}_z$, which based on the redefinition in \cite{dai2019information} we write in the case of pure states (product states), $I(\cdot)=\frac{1}{2}$,  mixed states (entangled states), $0<I(\cdot)<\frac{1}{2}$, and maximally mixed states (maximally entangled states), $I(\cdot)=0$.

The time-evolving reduced density operator of the atom $\hat{\overline{\rho}}_a^g(t)$ in Eq.~\eqref{eq:wigner1} is easily obtained by tracing \cite{gerry2005introductory,enriquez2010atomic,alber2003quantum,
kaye2007introduction,nielsen2011quantum} the time evolving AJC density matrix $\hat{\overline{\rho}}_{gr\alpha}(t)$ over the field mode states,  determined from Eq.~\eqref{eq:ajcev} according to
\begin{eqnarray}
\hat{\overline{\rho}}_a^g(t)&=&tr_f\hat{\overline{\rho}}_{gr\alpha}= tr_f\left(\mid\overline{\Psi}_{gr\alpha}(t)\rangle\langle\overline{\Psi}_{gr\alpha}(t)\mid\right)\nonumber\\
&=&\sum_{n=0}^\infty\Big[S_n^2\Big(\cos^2(\overline{R}_{gn}t)+\overline{c}_{gn}^2\sin^2(\overline{R}_{gn}t)\Big)|g\rangle\langle{g}|\nonumber\\&&
+i~S_n~S_{n-1}~\overline{s}_{gn-1}e^{-i\omega t}\sin(\overline{R}_{gn-1}t)\Big(\cos(\overline{R}_{gn}t)+i\overline{c}_{gn}\sin(\overline{R}_{gn}t)\Big)|g\rangle\langle e|\nonumber\\&&
-i~S_n~S_{n-1}~\overline{s}_{gn-1}e^{i\omega t}\sin(\overline{R}_{gn-1}t)\Big(\cos(\overline{R}_{gn}t)-i\overline{c}_{gn}\sin(\overline{R}_{gn}t)\Big)|e\rangle\langle g|\nonumber\\&&
+S_{n-1}^2~\overline{s}_{gn-1}^2\sin^2(\overline{R}_{gn-1}t)|e\rangle\langle e|\Big]~.\nonumber\\&&
\label{eq:redmat}
\end{eqnarray}
\end{subequations}

With reference to Eq.~\eqref{eq:ajcsumf}, at the standard resonance condition, $\delta=0$, i.e., $\omega_0=\omega$, the AJC Rabi frequency, $\overline{R}_{gn}$ and interaction parameters $\overline{c}_{gn},\,\overline{s}_{gn}$ in Eq.~\eqref{eq:ajcev} take the forms
\begin{eqnarray}
\overline{R}_{gn}=\lambda\sqrt{n+1+\xi^2}\quad;\quad\overline{c}_{gn}=\frac{\xi}{\sqrt{n+1+\xi^2}}\quad;\quad\overline{s}_{gn}=\frac{\sqrt{n+1}}{\sqrt{n+1+\xi^2}}~,
\label{eq:parares}
\end{eqnarray}
confirming the explicit dependence of $\overline{R}_{gn},\,\overline{c}_{gn},\,\overline{s}_{gn}$ on $\xi$. 

It therefore follows from Eq.\eqref{eq:parares}, zero difference frequency detuning, $\delta=0$, and consequently sum frequency, $\overline{\delta}=2\omega=2\xi\lambda$, will provide a clear picture of the effect of coupling strength, $\xi$, during the AJC atom-field interaction measured by the Wigner-Yanase skew information, $I\left(\hat{\overline{\rho}}_a^g(t),\,\hat{s}_z\right)$. In the following, incipiently Sec.~\ref{sec:copstr}, we evaluate the Wigner-Yanase skew information, $I\left(\hat{\overline{\rho}}_a^g(t),\,\hat{s}_z\right)$, defined in Eq.~\eqref{eq:wigner1}, specifically at $\delta=0;\quad\overline{\delta}=2\xi\lambda$, for different values of coupling strength, $\xi$ and provide the corresponding $I(\cdot)$ dynamical evolution curves.

It is also important to recall that when an ideal two-level atom interacts with a quantised optical field, the result is quantum collapses and revivals in atomic population inversion \cite{shore1993jaynes,shore_knight_1993,scully1997quantum,
satyanarayana1989ringing,gea1990collapse,gerry2005introductory,mayero2023photon,mayero2024squeezed}, where the revivals are an indicator of the nature of photon number distribution $P(n)$ Eq.~\eqref{eq:photonno} of the initial field mode inside a cavity. In this article, to have a clear visual of the non-classical nature of the AJC atom-field interaction measured by $I\left(\hat{\overline{\rho}}_a^g(t),\,\hat{s}_z\right)$, we introduce atomic population inversion, W(t), defined as the difference of time-evolving excited state probability, $\overline{P}_e(t)$, and time-evolving ground state probability, $\overline{P}_g(t)$,  according to
\begin{subequations}
\begin{eqnarray}
W(t)&=&\overline{P}_e(t)-\overline{P}_g(t)~,
\end{eqnarray}
which in terms of the time-evolving reduced density matrix of the atom, $\hat{\overline{\rho}}_a^g(t)$, explicitly defined in Eq.~\eqref{eq:redmat}, we write
\begin{eqnarray}
W(t)&=&tr(\hat{\sigma}_z\hat{\overline{\rho}}_a^g(t))~,\nonumber\\
&=&\sum_{n=0}^{\infty}\left[S_{n-1}^2\overline{s}_{gn-1}^2\sin^2(\overline{R}_{gn-1}t)-S_n^2\left(\cos^2(\overline{R}_{gn}t)+\overline{c}_{gn}^2\sin^2(\overline{R}_{gn}t)\right)\right]~.
\label{eq:redpop}
\end{eqnarray}
\end{subequations}
We use this expression in Eq.~\eqref{eq:redpop}, to evaluate W(t) and compare with that determined for $I\left(\hat{\overline{\rho}}_a^g(t),\,\hat{s}_z\right)$ at equal values of field intensity, $\mid\alpha\mid^2$, squeeze parameter, $r$, and coupling strength $\xi$ in Sec.~\ref{sec:copstr}.

Further investigation on the dynamics of I($\cdot$) during the AJC process set at AJC resonance condition $\overline{\delta}=2\xi\lambda;~\delta=0$ while varying field intensity, $\mid\alpha\mid^2$,  and squeeze parameter, $r$, is provided in Secs.~\ref{sec:inted},~\ref{sec:sqzp} respectively.
\subsection{Variation of coupling strength}
\label{sec:copstr}
\begin{figure}[H]
\centering
\subfloat[$\xi=0.0001$]{\label{fig:AJCS1}
\centering
\includegraphics[scale=0.85]{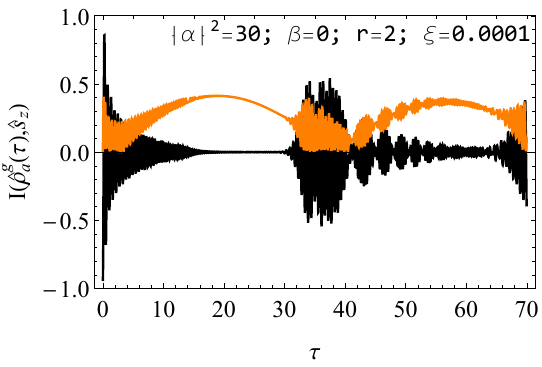}
}\hfill
\centering
\subfloat[$\xi=0.01$]{\label{fig:AJCS2}
\centering
\includegraphics[scale=0.85]{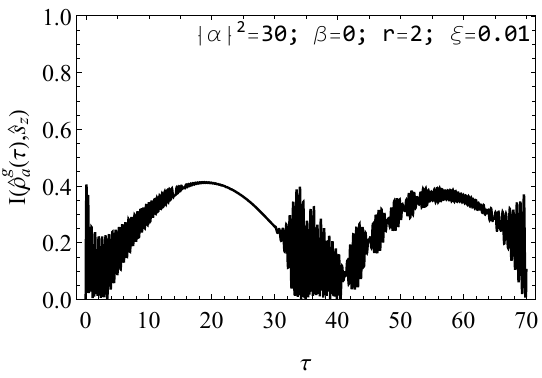}
}\\
\centering
\subfloat[$\xi=1.5$]{\label{fig:AJCS3}
\centering
\includegraphics[scale=0.85]{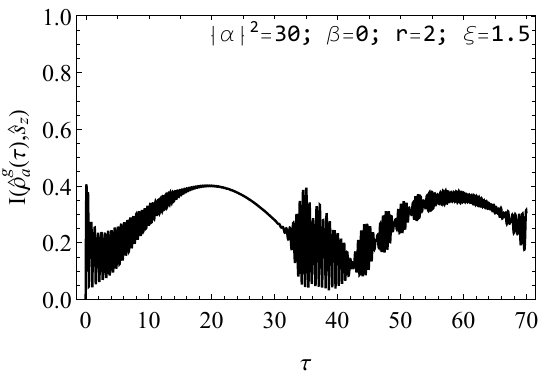}
}\hfill
\centering
\subfloat[$\xi=2$]{\label{fig:AJCS4}
\centering
\includegraphics[scale=0.85]{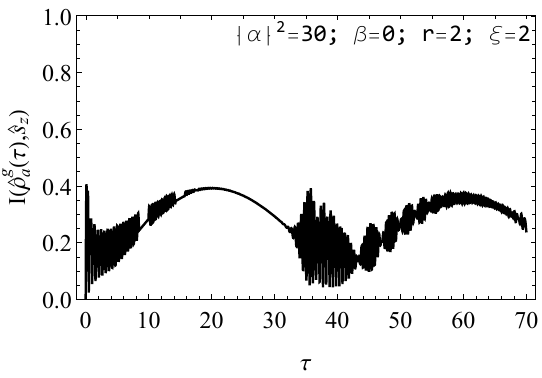}
}
\caption{Time evolution of Wigner-Yanase Skew Information $I\left(\hat{\overline{\rho}}_a^g(t),\,\hat{s}_z\right)$ at $\overline{\delta}=2\xi\lambda~(\beta=0, \delta=0)$, $r=2$, $\mid\alpha\mid^2=30$: Fig.~\eqref{fig:AJCS1}, W($\tau$)(black); I($\cdot$)(orange) at $\overline{\delta}=0.0002\lambda,~\xi=0.0001$, Fig.~\eqref{fig:AJCS2}, $\overline{\delta}=0.02\lambda,~\xi=0.01$, Fig.~\eqref{fig:AJCS3}, $\overline{\delta}=3\lambda,~\xi=1.5$ and Fig.~\eqref{fig:AJCS4} $\overline{\delta}=4\lambda,~\xi=2$.}
\label{fig:coup2024}
\end{figure}

The coupling strength parameter, $\xi$, is the ratio of the field mode frequency to coupling constant, $\omega:\lambda$, and it is a dimensionless quantity since $\omega,\,\lambda$ are defined in units of frequency, $s^{-1}$. We set field intensity, squeeze parameter, at arbitrary constant values of $\mid\alpha\mid^2=30,\,r=2$ and vary the coupling strength parameter, $\xi=0.0001,\,0.01,\,1.5,\,2$ in the plots of time evolution of the Wigner-Yanase skew information $I\left(\hat{\overline{\rho}}_a^g(\tau),\,\hat{s}_z\right)$  in Figs.~\eqref{fig:AJCS1},~\eqref{fig:AJCS2},~\eqref{fig:AJCS3},
~\eqref{fig:AJCS4} at $\beta=\frac{\delta}{\lambda}=0,~(\delta=0)$. Fig.~\eqref{fig:AJCS1}, provides an additional plot of time evolution of atomic population inversion,  W($\tau$) set at the same, $\mid\alpha\mid^2,\,r,\,\beta$ parameter values. In $I\left(\hat{\overline{\rho}}_a^g(\tau),\,\hat{s}_z\right),\,W(\tau)$, the temporal parameter, $\tau=\lambda t$ is the scaled time.
In the plots in Fig.~\ref{fig:coup2024} we see that:
\begin{enumerate}[label=(\roman{*})]
\item at the middle of collapse of atomic population inversion in Fig.~\ref{fig:AJCS1}, the system evolves to a low degree of mixedness, $I(\cdot)=0.4$, slightly below a pure state limit of $I(\cdot)=0.5$ and later evolves into a state of maximal mixedness, $I(\cdot)=0$, periodically similar to the collapse time, $\tau_c=\frac{1}{\sqrt{2}}$, in $I(\cdot),\,W(\tau)$ as a consequence of strongly coupled AJC atom-field interaction set at $\xi=0.0001$;
\item ringing revivals in $W(\tau),\,I(\cdot)$ are present in all plots in Fig.~\ref{fig:coup2024}. This is as a consequence of overlapping of additional peaks in the photon number distribution $P(n)$ Eq.~\eqref{eq:photonno} for all $r>1$, see \cite{satyanarayana1989ringing,subeesh2012effect,mayero2024squeezed,
moya1992interaction}. The additional peaks in the photon counting distribution are a result of an increased addition of squeezed photons in the coherent field mode, discussed earlier in \cite{satyanarayana1989ringing,moya1992interaction,subeesh2012effect,  mayero2024squeezed};
\item Fig.~\ref{fig:AJCS2} displays a similar form of time evolution of the Wigner-Yanase skew information, $I(\cdot)$, to that presented in Fig.~\ref{fig:AJCS1}, however notice a slight lift in $I(\cdot)$ at all time intervals above the maximally mixed state value, $I(\cdot)=0$, as a result of the reduced coupling, $\xi=0.01$. This can be visualised more clearly on Figs.~\ref{fig:AJCS3},~\ref{fig:AJCS4} where at respective weak coupling, $\xi=1.5,\,2$, the time evolution of $I(\cdot)$ is strictly mixed away from maximal mixedness $I(\cdot)=0$;
\item an decrease in the coupling strength from $\xi=0.01$ in Fig.~\ref{fig:AJCS2} to $\xi=1.5$ in Fig.~\ref{fig:AJCS3} and finally to $\xi=2$ in Fig.~\ref{fig:AJCS4}, results in delay in revivals leading to longer quiescent phases in the evolution of $I(\cdot)$ and; 
\item despite reduction in coupling strength parameter, $\xi\gg 0.0001$,  resulting into longer collapse period in $I(\cdot)$ Figs.~\ref{fig:AJCS3},~\ref{fig:AJCS4}, the atom-field system is maintained in a mixed state range, $0<I(\cdot)<0.5$, at all time intervals as expected when an initial squeezed field-mode state is considered \cite{mayero2024squeezed,moya1992interaction}.
\end{enumerate}
\subsection{Variation of field intensity}
\label{sec:inted}
\begin{figure}[H]
\centering
\subfloat[$\mid\alpha\mid^2=10$]{\label{fig:AJCF1}
\centering
\includegraphics[scale=0.85]{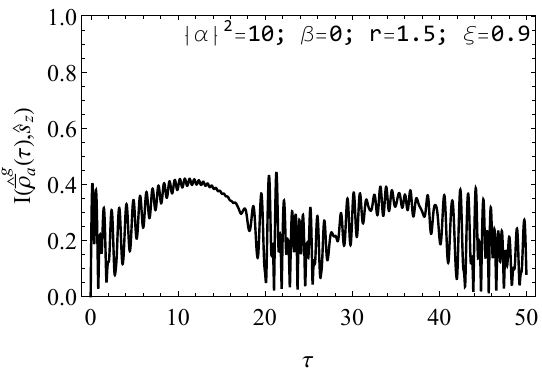}
}\hfill
\centering
\subfloat[$\mid\alpha\mid^2=20$]{\label{fig:AJCF2}
\centering
\includegraphics[scale=0.85]{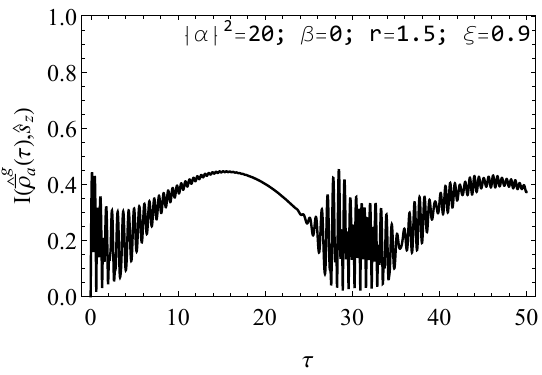}
}\\
\centering
\subfloat[$\mid\alpha\mid^2=30$]{\label{fig:AJCF3}
\centering
\includegraphics[scale=0.85]{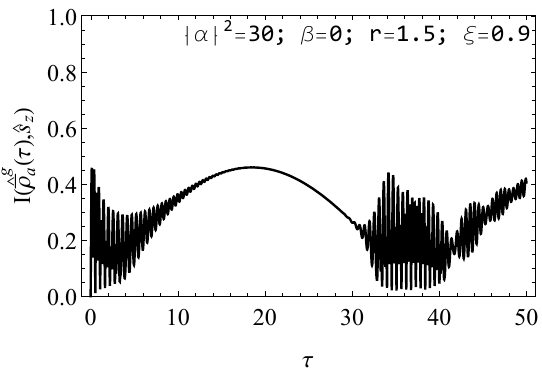}
}\hfill
\centering
\subfloat[$\mid\alpha\mid^2=40$]{\label{fig:AJCF4}
\centering
\includegraphics[scale=0.85]{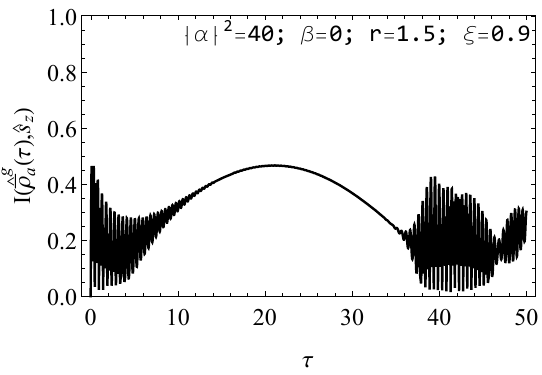}
}
\caption{Time evolution of Wigner-Yanase Skew Information $I\left(\hat{\overline{\rho}}_a^g(\tau),\hat{s}_z\right)$ at $\overline{\delta}=2\xi\lambda=1.8\lambda~(\xi=0.9),\,r=1.5$: Fig.~\eqref{fig:AJCF1}, $\mid\alpha\mid^2=10$, Fig.~\eqref{fig:AJCF2}, $\mid\alpha\mid^2=20$, Fig.~\eqref{fig:AJCF3}, $\mid\alpha\mid^2=30$ and Fig.~\eqref{fig:AJCF4}, $\mid\alpha\mid^2=40$.}
\label{fig:intfigs}
\end{figure}

In this section we investigate the effect of field intensity, $\mid\alpha\mid^2$, on the dynamics of the Wigner-Yanase skew information, $I\left(\hat{\overline{\rho}}_a^g(\tau),\hat{s}_z\right)$, while keeping difference frequency detuning, $\delta$,  squeeze parameter, $r$,  coupling strength parameter, $\xi$, fixed at $\delta=\omega_0-\omega=0;~\beta=0$, $r=1.5$, $\xi=0.9$. We arbitrarily set the field intensity at, $\mid\alpha\mid^2=10,\,20,\,30,\,40$, and provide the $I(\cdot)$ dynamical evolution plots in Figs.~\ref{fig:AJCF1},~\ref{fig:AJCF2},~\ref{fig:AJCF3} and~\ref{fig:AJCF4} respectively.
With reference to the plots in Fig.~\eqref{fig:intfigs}:
\begin{enumerate}[label=(\roman{*})]
\item an increase in field intensity results in more rapid oscillations in $I(\cdot)$ as visualised in Figs.~\ref{fig:AJCF1},~\ref{fig:AJCF2},~\ref{fig:AJCF3},
~\ref{fig:AJCF4} in the order of increasing field intensity, $\mid\alpha\mid^2$;
\item longer revival times, $\tau_R$, in the time evolution of $I(\cdot)$ is evident in Fig.~\eqref{fig:AJCF2} compared to Fig.~\eqref{fig:AJCF1}; Fig.~\eqref{fig:AJCF3} in comparison to Figs.~\eqref{fig:AJCF1},~\eqref{fig:AJCF2}; 
Fig.~\eqref{fig:AJCF4} compared to Figs.~\eqref{fig:AJCF1},~\eqref{fig:AJCF2},~\eqref{fig:AJCF3}    again for every rise in mean photon number and;
\item the weak coupling set at $\xi=0.9$, is not affected by variation in mean photon number, i.e., it is not weakened further or made stronger. Notice the almost constant upward shift above the maximally mixed state limit, $I(\cdot)=0$ \cite{dai2019information} for all variations in $\mid\alpha\mid^2$.
\end{enumerate}

\subsection{Variation of squeeze parameter}
\label{sec:sqzp}
\begin{figure}[H]
\centering
\subfloat[$r=0.5$]{\label{fig:AJCR1}
\centering
\includegraphics[scale=0.85]{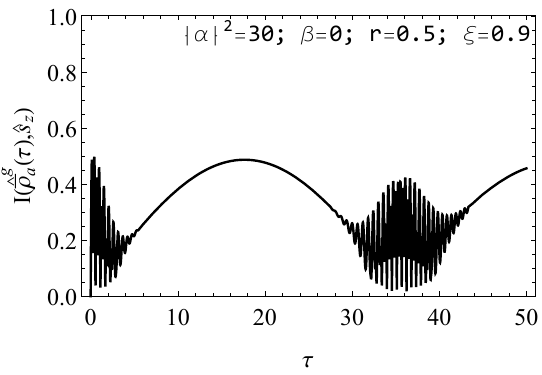}
}\hfill
\centering
\subfloat[$r=1$]{\label{fig:AJCR2}
\centering
\includegraphics[scale=0.85]{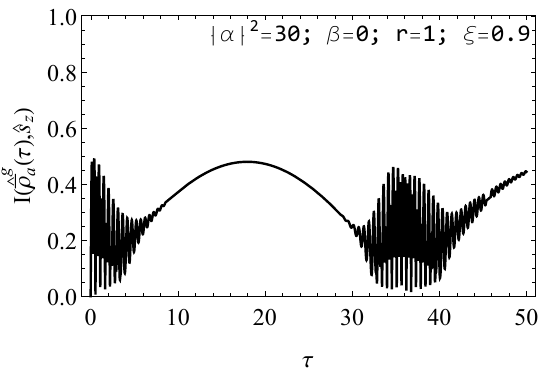}
}\\
\centering
\subfloat[$r=1.5$]{\label{fig:AJCR3}
\centering
\includegraphics[scale=0.85]{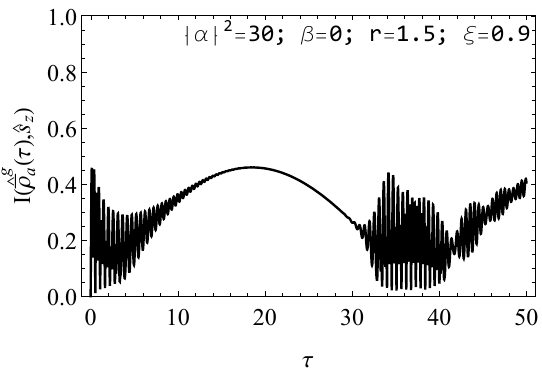}
}\hfill
\centering
\subfloat[$r=2$]{\label{fig:AJCR4}
\centering
\includegraphics[scale=0.85]{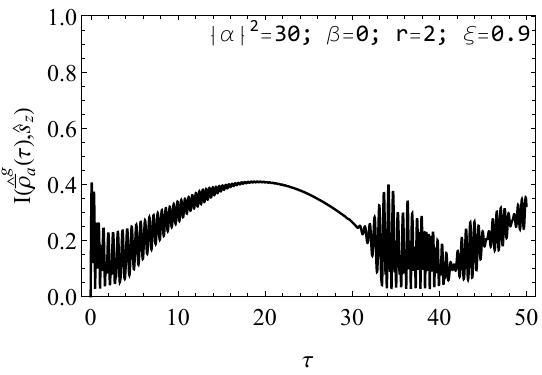}
}
\caption{Time evolution of Wigner-Yanase Skew Information $I\left(\hat{\overline{\rho}}_a^g(\tau),\hat{s}_z\right)$ at $\overline{\delta}=2\xi\lambda=1.8\lambda~(\xi=0.9),\,\mid\alpha\mid^2=30$: Fig.~\eqref{fig:AJCR1}, $r=0.5$, Fig.~\eqref{fig:AJCR2}, $r=1$, Fig.~\eqref{fig:AJCR3}, $r=1.5$ and Fig.~\eqref{fig:AJCR4}, $r=2$.}
\label{fig:squzfigs}
\end{figure}

We evaluate the Wigner-Yanase skew information in Eq.~\eqref{eq:wigner1} at fixed field intensity, $\mid\alpha\mid^2=30$, coupling strength, $\xi=0.9$, difference frequency detuning, $\delta=0, \beta=0$ while varying upwards the squeeze parameter, $r=0.5,\,1,\,1.5,\,2$ and provide the corresponding plots in Figs.~\ref{fig:AJCR1},~\ref{fig:AJCR2},~\ref{fig:AJCR3} and~\ref{fig:AJCR4}.

From the plots in Fig.~\ref{fig:squzfigs}, we conclude the following:
\begin{enumerate}[label=(\roman{*})]
\item the effect of weak atom-field coupling $\xi=0.9$ is evident,  marked by a slight vertical shift in the time evolution of the Wigner-Yanase skew information above $I(\cdot)=0$ in Figs.~\ref{fig:AJCR1},~\ref{fig:AJCR2},~\ref{fig:AJCR3},
~\ref{fig:AJCR4} and it is unaffected by upward variations in squeeze parameter, $r$;
\item in the example in Fig.~\ref{fig:AJCR1} set at low squeeze parameter, $r=0.5$, the evolution tends to attain the coherent pure state value, $I(\cdot)\simeq\frac{1}{2}$ \cite{dai2019information,dai2020atomic} at the middle of collapse of $W(\tau)$ (see example in Fig.~\ref{fig:AJCS1}). Also notice that there is no ringing after the revival phase, $\tau>40$ and;
\item as the squeeze parameter is increased (see Figs.~\ref{fig:AJCR2},~\ref{fig:AJCR3},~\ref{fig:AJCR4}), the oscillations in $I(\cdot)$ becomes more rapid, the peak value in $I(\cdot)$ reduces with every increase in $r$ indicating diminished tendency of the atom-field states to evolve to pure state (see Figs.~\ref{fig:AJCR3},~\ref{fig:AJCR4}) and ringing revivals \cite{moya1992interaction,mayero2024squeezed} are evident at $\tau>40$.
\end{enumerate}
\section{Conclusion}
\label{sec:conc}
In this study, we have analysed atomic non-classicality measured by the Wigner-Yanase skew information during the AJC interaction when an initial squeezed coherent field mode is considered. It is clear that reducing atom-field coupling, $\xi=\frac{\omega}{\lambda}$, which can also be viewed as an upward variation in sum frequency, $\overline{\delta}=2\xi\lambda$, preserves the set field-mode squeezing for a longer period of time. The net effect as visualised in the provided examples, is delay in revival of atomic inversion coupled with a decrease in degree of entanglement, i.e., the atom-field states remain mixed as time develops and not maximally mixed. What is more, is that increasing the field intensity results in improved oscillations in the dynamics of the Wigner-Yanase skew information and a simultaneous delay in revival of atomic inversion. In this set-up, we see that for a fixed field-mode squeezing, $r$, upping the field intensity, the atom-field states tend to evolve to an approximate pure state, $I(\cdot)\simeq \frac{1}{2}$, at the middle of collapse of atomic population inversion similar to when a coherent field mode is considered \cite{dai2020atomic}.  This is because the coherent part, $\mid\alpha\mid^2$, of the squeezed coherent field-mode state dominates the squeezed part, $\sinh(r)$. The ringing revivals is however not affected as provided in the examples. Finally, when we increase the squeeze parameter, $r$, the oscillations in the temporal evolution of the Wigner-Yanase skew information become more rapid, occur in the mixed state range $0<I(\cdot)<\frac{1}{2}$ and  develop ringing revivals basically due to an increased number of squeezed photons in the coherent field.
\section*{Acknowledgement}

I thank Tom Mboya University in Kenya, for availing the required infrastructure to carry out this study.

\section*{Data availability}

All data required is available in this manuscript. 

\section*{Disclosure statement}

The author declares no conflict of interest towards publication of this work.

\section*{Notes on contributors}

The author wrote and analysed each section of this manuscript as presented.

\bigbreak
\def\bibsection{}
\centerline{\textbf{REFERENCES}}
\bigbreak

\end{document}